\documentclass[apjl]{emulateapj}

\usepackage{amsmath,amssymb}
\usepackage{natbib}
\usepackage{epsfig,graphics}
\usepackage{epstopdf} 
\usepackage{ulem}

\newcommand{\lsim}{\raise0.3ex\hbox{$<$}\kern-0.75em{\lower0.65ex\hbox{$\sim$}}}
\newcommand{\gsim}{\raise0.3ex\hbox{$>$}\kern-0.75em{\lower0.65ex\hbox{$\sim$}}}

\shorttitle{Eclipse Measurements of WASP-18b}
\shortauthors{Sheppard et al.}

\begin{document}

\title{Evidence for a Dayside Thermal Inversion and High Metallicity for the Hot Jupiter WASP-18b}

\author{Kyle B. Sheppard\altaffilmark{1,2,5}, Avi M. Mandell\altaffilmark{2}, Patrick Tamburo\altaffilmark{2,3}, Siddharth Gandhi\altaffilmark{4}, Arazi Pinhas\altaffilmark{4}, Nikku Madhusudhan\altaffilmark{4}, Drake Deming\altaffilmark{1} 
}
\altaffiltext{1}{Department of Astronomy, University of Maryland, College Park, MD 20742, USA}
\altaffiltext{2}{Solar System Exploration Division, NASA's Goddard Space Flight Center, Greenbelt, MD 20771, USA}
\altaffiltext{3}{Department of Astronomy, Boston University, Boston, MA 02215, USA}
\altaffiltext{4}{Institute of Astronomy, University of Cambridge, UK}
\altaffiltext{5}{Corresponding Email: kshep23@umd.edu}

\slugcomment{Draft Version; not for circulation}

\begin{abstract}

 We find evidence for a strong thermal inversion in the dayside atmosphere of the highly irradiated hot Jupiter WASP-18b (T$_{eq}=2411K$, $M=10.3M_{J}$) based on emission spectroscopy from Hubble Space Telescope secondary eclipse observations and \textit{Spitzer} eclipse photometry. We demonstrate a lack of water vapor in either absorption or emission at 1.4\,$\mu$m. However, we infer emission at 4.5\,$\mu$m and absorption at 1.6\,$\mu$m that we attribute to CO, as well as a non-detection of all other relevant species (e.g., TiO, VO). The most probable atmospheric retrieval solution indicates a C/O ratio of 1 and a high metallicity (C/H=$283^{+395}_{-138}\times$ solar). The derived composition and T/P profile suggest that WASP-18b is the first example of both a planet with a non-oxide driven thermal inversion and a planet with an atmospheric metallicity inconsistent with that predicted for Jupiter-mass planets at $>2\sigma$. Future observations are necessary to confirm the unusual planetary properties implied by these results.

\end{abstract}


\section{Introduction}

Hot Jupiters have been vital in revealing the structural and atmospheric diversity of gas-rich planets \citep[see recent reviews by][]{crossfield,madhu2016,demseg}. Since they are exposed to extreme conditions and relatively easy to observe through transit and eclipse spectroscopy, hot Jupiters provide a window into a unique part of parameter space, allowing us to better understand both atmospheric physics and planetary structure.

An outstanding question that has emerged for highly irradiated planets is the presence and origin of stratospheric thermal inversions, which have been detected in several extremely irradiated hot Jupiters \citep{Haynes:2015cf,Evans:2017dp}. \citet{Hubeny:2003eb} predicted that thermal inversions in highly-irradiated atmospheres would be caused by the presence of optical absorbers (e.g. TiO and VO) high in the atmosphere, but there may be other causes such as insufficient cooling \citep{Molliere:2015} or sulfur-based aerosols \citep{zahnle}.

Constraints on the structure and composition of exoplanetary atmospheres allow us to test, refine, and generalize planetary formation models. Volatile ices are expected to play an important role in planet formation; thus a constraint on the composition of a hot planet's atmosphere gives us insight on how and where it was formed \citep{Oberg:2011je,Madhusudhan:2014}. In our Solar System there is an inverse mass vs. atmospheric metallicity relationship, and whether or not it extends to exoplanets is informative to planetary formation and migration models. There is some evidence that the trend holds \citep{Kreidberg:2014hi}, however that parameter space is not yet sufficiently populated to enable firm conclusions. 

In this paper we use Hubble Space Telescope (HST) spectroscopy and {\it Spitzer}/IRAC photometry of secondary eclipses to explore the thermal structure and composition of the dayside atmosphere of WASP-18b, an extremely hot ($T_{eq} = 2411$K) and massive ($M=10.3M_{J}$) hot Jupiter orbiting an F-type star with an orbital period of less than one day \citep{Hellier:2009}.

\section{Observations}

We used Wide Field Camera-3 (WFC3) observations of five secondary eclipses of WASP-18b from the HST Treasury survey by Bean et al. (Program ID 13467). WFC3 obtains low resolution slitless spectroscopy from 1.1 to 1.7\,$\mu$m using the G141 grism (R=130), as well as an image for wavelength calibration using the F140W filter. Grism observations were taken in spatial scan mode \citep{Deming:2013ge} with a forward-reverse cadence \citep{Kreidberg:2014hi}. The first three visits, taken between April-June 2014, are single eclipse events. Visit 4, taken in August 2014, contains two eclipses in an orbital phase curve, and we extract those eclipses and analyze them separately.

We also re-analyze two eclipse observations of WASP-18b taken in the 3.6\,$\mu$m and 4.5\,$\mu$m channels of the \textit{Spitzer Space Telescope}'s IRAC instrument (Program ID 60185). The 3.6 \,$\mu$m observation was performed on 2010 January 23, while the 4.5 \,$\mu$m observation was taken 2010 August 23. Both observations were taken using an exposure time of 0.36s in subarray mode, and were first analyzed in \cite{Maxted:2013ey}. 

\section{\textit{HST} Data Analysis}

Our grism spectroscopy analysis utilized HST ``ima'' data files.  We separated the data by scan direction, removed background flux, and corrected for cosmic rays and bad pixels. We removed background flux via the ``difference frames'' method outlined in the appendix of \citet{Deming:2013ge}, and set the aperture to maximize the amount of source photons in our analysis. The end result is two reduced light curves - one forward scan and one reverse scan - for each eclipse, which we analyze separately.

The F140W photometric image determines the location of the zero-point, which we used to assign a wavelength to each column. We confirmed the wavelengths by fitting a T=6400K, log\,$g$=4.3, [Fe/H]=0.1 ATLAS stellar spectrum  \citep{Castelli:2004ti}, multiplied by the grism sensitivity curve, to an observed in-eclipse spectrum. 

\subsection{Light Curve Analysis}

Empirical methods are necessary to correct for non-astrophysical systematic effects in WFC3 spectroscopy \citep{Berta:2012p7117,Haynes:2015cf}. Correction methodology is especially important in emission spectroscopy, where the magnitude of systematic effects can be greater than the eclipse depth \citep{Kreidberg:2014hi}. We thereby combined two strategies: initial removal of systematic trends using parametric marginalization \citep{gibson, wakeford}, and further detrending by subtraction of scaled band-integrated residuals from wavelength bins \citep{Mandell:2013hp, Haynes:2015cf}. Our method accounts for uncertainty in instrument model selection, and residuals from the band-integrated analysis allow us to utilize the normally excluded first orbit of each HST data set in our spectroscopic analysis.

Fitting a band-integrated light curve provides residuals that we use to remove unidentified systematics from the spectrally resolved light curves. We calculate the HST phase (parameter for ramp and HST breathing), planetary phase (parameter for visit-long slope), and a wavelength shift derived by cross correlating each spectrum with the last spectrum for the visit (parameter for jitter) for each exposure in a time series. The grid of systematic models comprises a combination of a linear planetary phase correction and up to four powers of HST phase and wavelength shift. These models are then multiplied by a \citet{Mandel:2002bb} eclipse model. We simultaneously fit for the eclipse depth, all systematic coefficients, and - for two light curves with ingress and egress points - the center of eclipse time. All other system parameters are fixed to literature values. 
  
\begin{figure}[t]
\centering
{
\includegraphics[width=170mm]{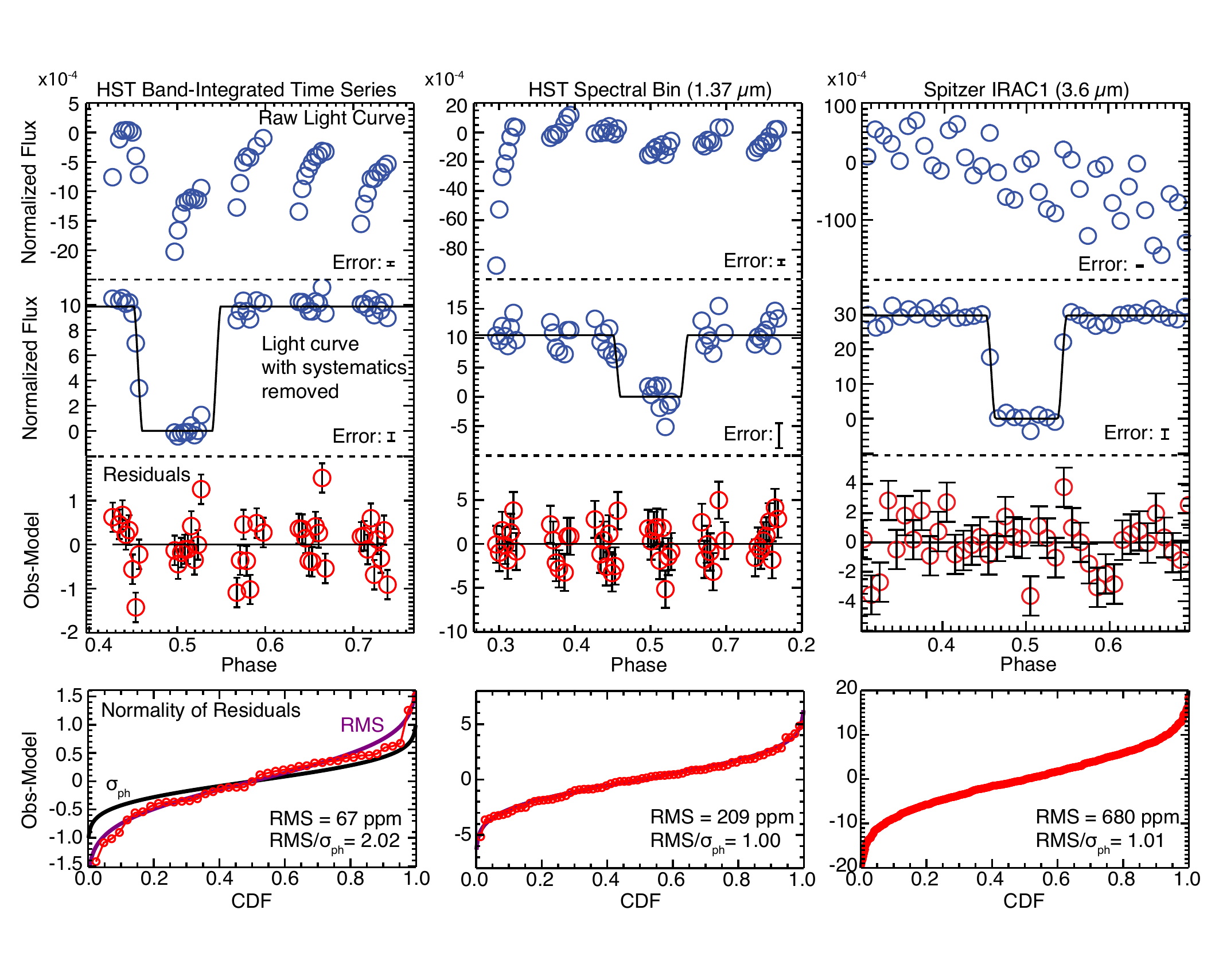}
\caption{An example of the detrending process for an HST band-integrated light curve (left), a light curve for an HST spectral bin (middle), and a \textit{Spitzer}/IRAC photometry light curve (binned for clarity). The HST band-integrated results fall within $1.3-5.5\times$ the photon noise limit, while both the HST spectral bins and the \textit{Spitzer} data typically achieve close-to-photon-limited results. The bottom row compares the cumulative distribution function (CDF) of the residuals to that of a Gaussian with dispersion equal to the photon noise. Good agreement is obtained for the HST spectral and Spitzer residuals, while excess scatter is observed for the HST band-integrated residuals. For the latter, the CDF of a Gaussian with dispersion equal to the residual RMS is also plotted for comparison.}
\label{fig:data}
}
\end{figure}

We use a Levenberg-Markwardt (L-M) least squares minimization algorithm \citep{markwardt} to determine the parameter values, since \citet{wakeford} found it to agree with the more computationally expensive Monte Carlo Markov Chain analysis to within 10\%. An example band-integrated light curve with systematic effects removed using the best-fitting model is shown in the leftmost panel of Figure~\ref{fig:data}. The scatter (RMS) of the residuals of the band-integrated curves ranges from 1.3-5.5$\times$ the photon noise, indicating that there is excess noise beyond the photon limit present. Excess noise in the band-integrated curves is also shown by comparisons of the cumulative distributions of residuals with those of a photon-limited Gaussian (see bottom-left panel of Figure 1). However, the structure of this excess noise does not change with wavelength, allowing for its removal from the corresponding spectral light curves.

To derive the emission spectrum, we bin the exposures in wavelength between the steep edges of the grism response and fit these spectrally resolved light curves. We remove wavelength-dependent systematics by fitting each spectral bin separately in a process that mimics the band-integrated process, with three exceptions. First, the eclipse mid-time is now fixed to the value determined by the band-integrated analysis. Second, it is possible that shifts on the detector are wavelength-dependent, so the jitter parameter is recalculated for each wavelength bin using only that portion of the spectrum in the cross-correlation procedure. Third, each systematic model now incorporates the residuals from the band-integrated fit of the same model as a decorrelation variable. The amplitude of the residuals is a free paramter, although the shape is assumed to be constant in wavelength. This removes any remaining wavelength-independent trends in the data. An example result of a reduced spectral bin light curve is shown in the central panel of Figure~\ref{fig:data}.

\begin{figure}[t]
\centering
{
\includegraphics[width=170mm]{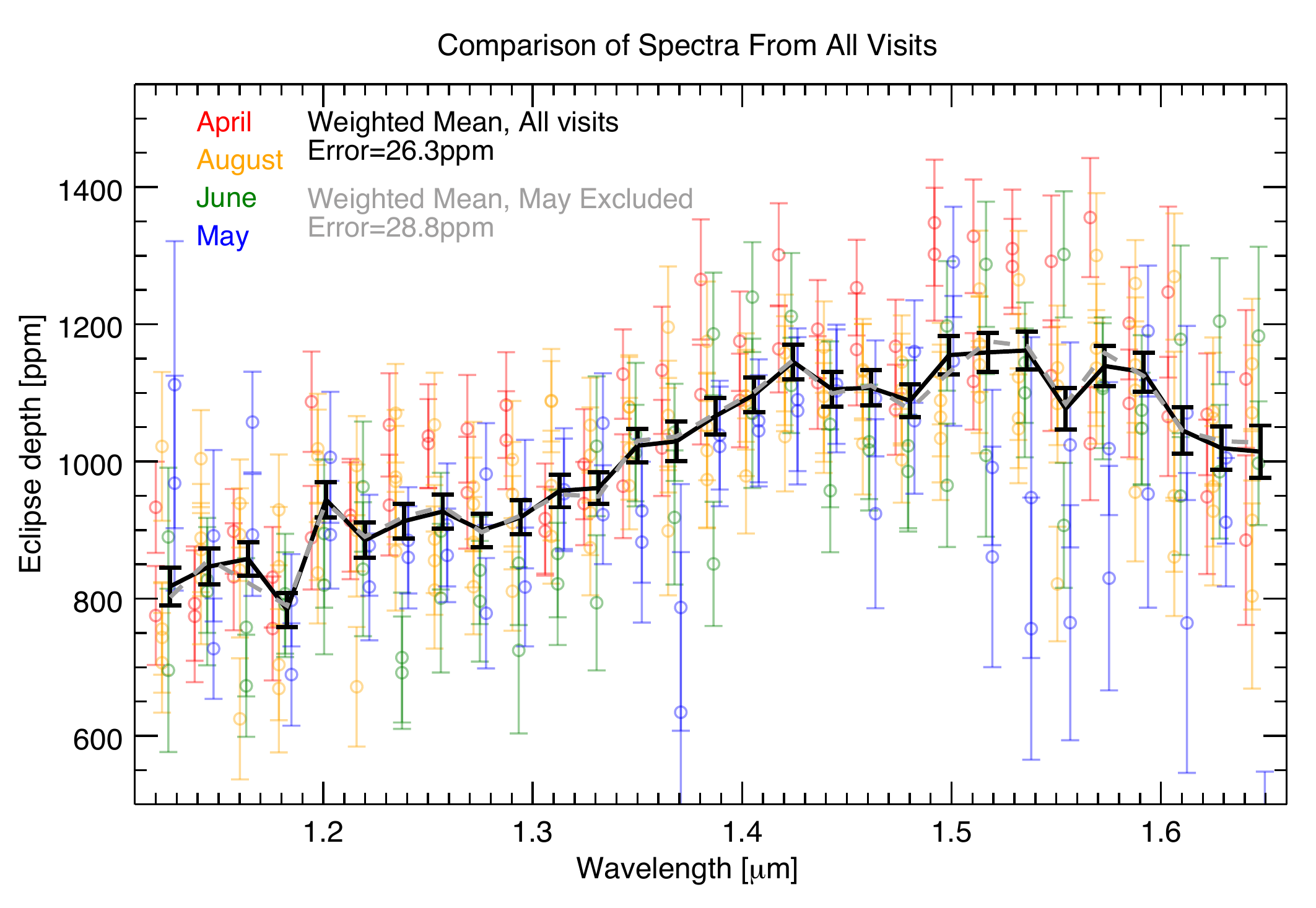}
\caption{Spectra for all of the HST visits, horizontally offset for clarity, with the weighted mean overplotted. Depths from both the forward and reverse scan light curves are plotted for each eclipse. The May data receives a low weight due to the large uncertainties, and therefore does not impact the results beyond the individual uncertainties, as shown by the dashed grey line. Values for the individual data points are available from the authors upon request.}
\label{fig:visits}
}
\end{figure}

Finally, eclipse depths from the multiple visits are combined via an inverse-variance weighted mean, giving the emission spectrum for WASP-18b. The spectra for all visits are shown in Figure~\ref{fig:visits}. 

The average RMS of the systematic-reduced spectroscopic light curves is 1.04$\times$ the photon noise and the median RMS is 0.97$\times$ the photon noise, indicating that shot noise is typically the dominant error source. The close agreement between the cumulative distributions of residuals and those of a Gaussian with a width determined by the photon noise provides further evidence that the analysis achieved photon-limited results for the vast majority of spectral curves (see bottom-center panel of Figure 1).The remaining spectral curves have residuals with an RMS greater than 1.5$\times$ the photon limit, indicating that excess noise is present. These only constitute 6\% of all spectral bins, and every one is from the single eclipse observation taken in May. We explored removing the May dataset due to this increased noise, but the exclusion of these data did not affect the variance-weighted spectrum, and we chose to include this visit in subsequent analyses. Figure~\ref{fig:visits} contains the emission spectra from every visit, demonstrating the consistency of the structure of the spectrum. Our analysis routine finds that the outlier depths from the May visit have very high errors due to the presence of correlated noise, and so they contribute very little to the weighted spectrum.

To further check our methodology, we reanalyzed published emission spectra for WASP-43b, WASP-103b, and WASP-121b. We find an agreement to the published spectra, with a mean point-by-point variation (difference / uncertainty) of 89\%, 23\%, and 50\% for the three data sets, respectively, demonstrating the consistency of our analysis pipeline with those published by other authors.

\section{\textit{Spitzer} Re-analysis}

Spitzer secondary eclipse measurements of WASP-18b were reported by \citet{Maxted:2013ey}, and we have re-analyzed key portions of those data.  We confine our re-analysis to the 3.6 and 4.5\,$\mu$m bands, because the instrumental systematic errors are greatest in those bands, and there are new methods to correct those systematics. 

We use an updated version \citep{Tamburo:2017} of the Pixel-Level Decorrelation framework \citep{Deming:2015fa}. Our photometry uses 11 different circular aperture sizes (with radii ranging from 1.6 to 3.5 pixels). We decorrelate the instrumental systematics while simultaneously fitting for the eclipse depth, using binned data, as advocated by \citet{Deming:2015fa} and \citet{Kammer:2015dn}. The fitting code selects the optimal aperture and bin size, and obtains an initial estimate of the eclipse depth and the pixel basis vector coefficients using linear regression.  We then implement an MCMC procedure \citep{Ford:2005p8180} to explore parameter space, refine the best-fit values, and determine the errors. At each step, we allow the central phase, orbital inclination, and eclipse depth to vary,  but lock all other orbital parameters to the values used in the WFC3 analysis. We also vary the multiplicative coefficients of our basis pixels (see \citealp{Deming:2015fa}) and visit-long quadratic temporal baseline coefficients at every step.  Our best fits use aperture radii of 2.0 and 2.5 pixels, and bin sizes of 76 and 116 points at 3.6 and 4.5\,$\mu$m, respectively. The scatter in the binned data, after removal of the best-fit eclipse, is 1.01 and 0.95$\times$ the photon noise at 3.6 and 4.5\,$\mu$m, respectively, those ratios being statistically indistinguishable from unity.

We ran three chains of 500,000 steps for both bands, confirming their convergence through the Gelman-Rubin statistic \citep{Gelman:1992ht}. We combine all chains of eclipse depth into a unified posterior distribution for each band, and fit a Gaussian to this distribution to determine the error on eclipse depth. Our results are included in Table 1, and exhibit excellent agreement with \citet{Maxted:2013ey}, but with smaller errors. 

\section{Atmospheric Retrieval}

We use our WFC3 spectrum along with the Spitzer and ground-based Ks band photometry to constrain the composition and temperature structure of the dayside atmosphere of WASP-18b. We use the HyDRA retrieval code \citep{Gandhi:2017_retrieval}, which comprises a thermal emission model of an atmosphere coupled with a nested sampling algorithm for Bayesian inference and parameter estimation. The forward model, based on standard prescriptions for retrieval \citep{Madhusudhan:2009p4072,madhu2011}, computes line-by-line radiative transfer in a plane parallel atmosphere under the assumptions of hydrostatic equilibrium and local thermodynamic equilibrium. The pressure-temperature ($P$-$T$) profile and chemical compositions are free parameters in the model. 

The model includes 14 free parameters. For the $P$-$T$ profile, we use the parametrisation of \citep{Madhusudhan:2009p4072} which involves six free parameters. The atmosphere comprises 100 layers equally spaced in log-pressure between $10^{-6}$ bar and $10^{2}$ bar. For the atmospheric composition we consider several species expected to be prevalent in very hot Jupiter atmospheres and with significant opacity in the observed spectral range \citep{madhu2012,Moses:2013jh,venot2015}. This includes H$_2$O, CO, CH$_4$, CO$_2$, HCN, C$_2$H$_2$, TiO, and VO. The uniform mixing ratio of each species are free parameters in the model. We assume an H$_2$/He rich atmosphere with a solar He/H$_2$ ratio of 0.17. We consider line absorption from each of these species and collision-induced opacity from H$_2$-H$_2$ and H$_2$-He. The sources of opacity data are described in \citet{Gandhi:2017ty}; the molecular linelists are primarily from EXOMOL \citep{Tennyson:2016ft} and HITEMP \citep{Rothman:2010p4751}, and the CIA opacities are from \citet{richard2012}. The retrieval explores model parameter space with Bayesian nested sampling using the MultiNest code via the Python wrapper, PyMultiNest \citep{skilling2004, feroz2013, buchner2014}. We sample the multi-dimensional parameter space using 4,000 live points for a total of more than one million model evaluations.

\begin{figure}[t]
\centering
{
\includegraphics[width=170mm]{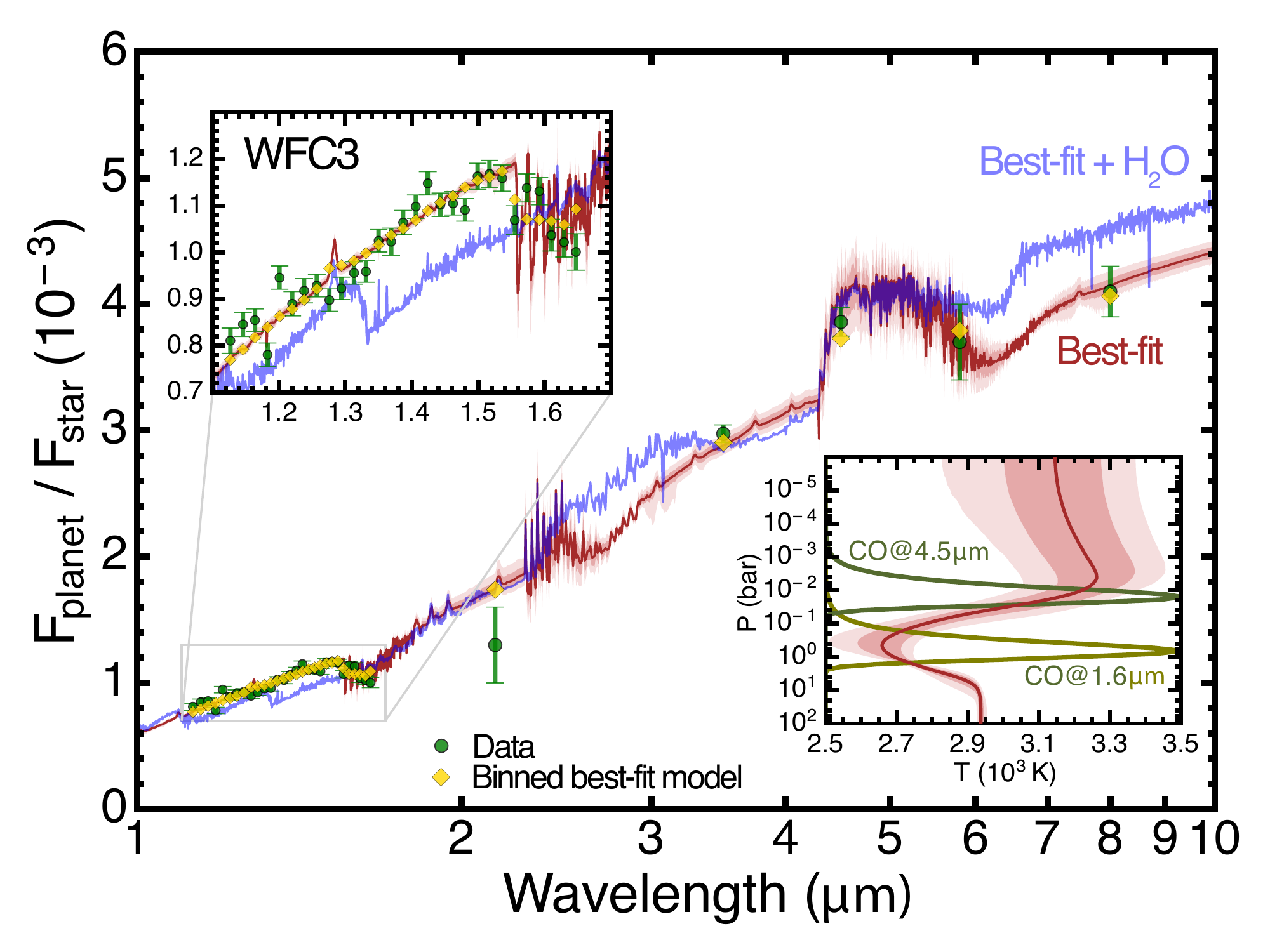}
\caption{ Observed spectrum and retrieved solutions. Our WFC3 and Spitzer data are shown in green. The median retrieved spectrum, with the uncertainity envelopes, is shown in red. The binned median model, in yellow, with $\chi^{2}_{red}=3.67$ is an unambiguously better fit than a blackbody ($\chi^{2}_{red}=15.2$). A fiducial model with solar-abundance H$_2$O absorption is shown in blue to demonstrate the lack of an H$_2$O feature in the data. The results favor a thermal inversion, and the only spectral features detected are those of CO at 1.6 and 4.5$\;\mu$m. The retrieved P-T profile with error contours is shown in the lower-right inset along with normalized contributions functions at 1.6 and 4.5 $\mu$m.}
\label{fig:spectrum}
}
\end{figure}

Our best-fit retrieval requires a strong thermal inversion in the dayside atmosphere. The bottom inset of Figure~\ref{fig:spectrum} shows the retrieved $P$-$T$ profile with confidence contours, indicating an upper atmospheric temperature increase. The requirement of a thermal inversion is guided by the strong emission inferred in the 4.5 $\mu$m Spitzer IRAC band, with a brightness temperature of 3100$\pm$50\,K, which is significantly higher than the rest of the data. This can be explained by the presence of a thermal inversion in the atmosphere along with the presence of either CO or CO$_2$, which both exhibit pronounced spectral features in the 4.5\,$\mu$m band \citep{Burrows:2007dw,Fortney:2008p81, madhu2010}. We break this degeneracy by requiring that CO$_2$ be less than H$_2$O as expected for hot Jupiter atmospheres \citep{madhu2012,heng_lyons_2016}.  Another subtlety is the apparent minor trough near $\sim1.6$\,$\mu$m, which we attribute to CO absorption below the inversion layer ($\sim$ 1-10 bar), where temperature decreases outward. Emission in the 4.5\,$\mu$m band is due to CO in the 0.001 - 0.1 bar range which contains the thermal inversion. As part of the nested sampling analysis, we compute the Bayesian evidence value for our retrieved spectrum. By comparing this value with that obtained for a model without a thermal inversion, we conclude that a thermal inversion is favored at the 6.3$\sigma$ significance level. Similarly, comparison to a model lacking CO implies that the presence of CO is favored at the 6.1$\sigma$ level. Interestingly, the transition point of the inversion occurs at 0.1 bar which is characteristic of all planets in the Solar System with inversions as well as models of hot Jupiters \citep{Madhusudhan:2009p4072,robinson2014}. 

\begin{figure}[t]
\centering
{
\includegraphics[width=170mm]{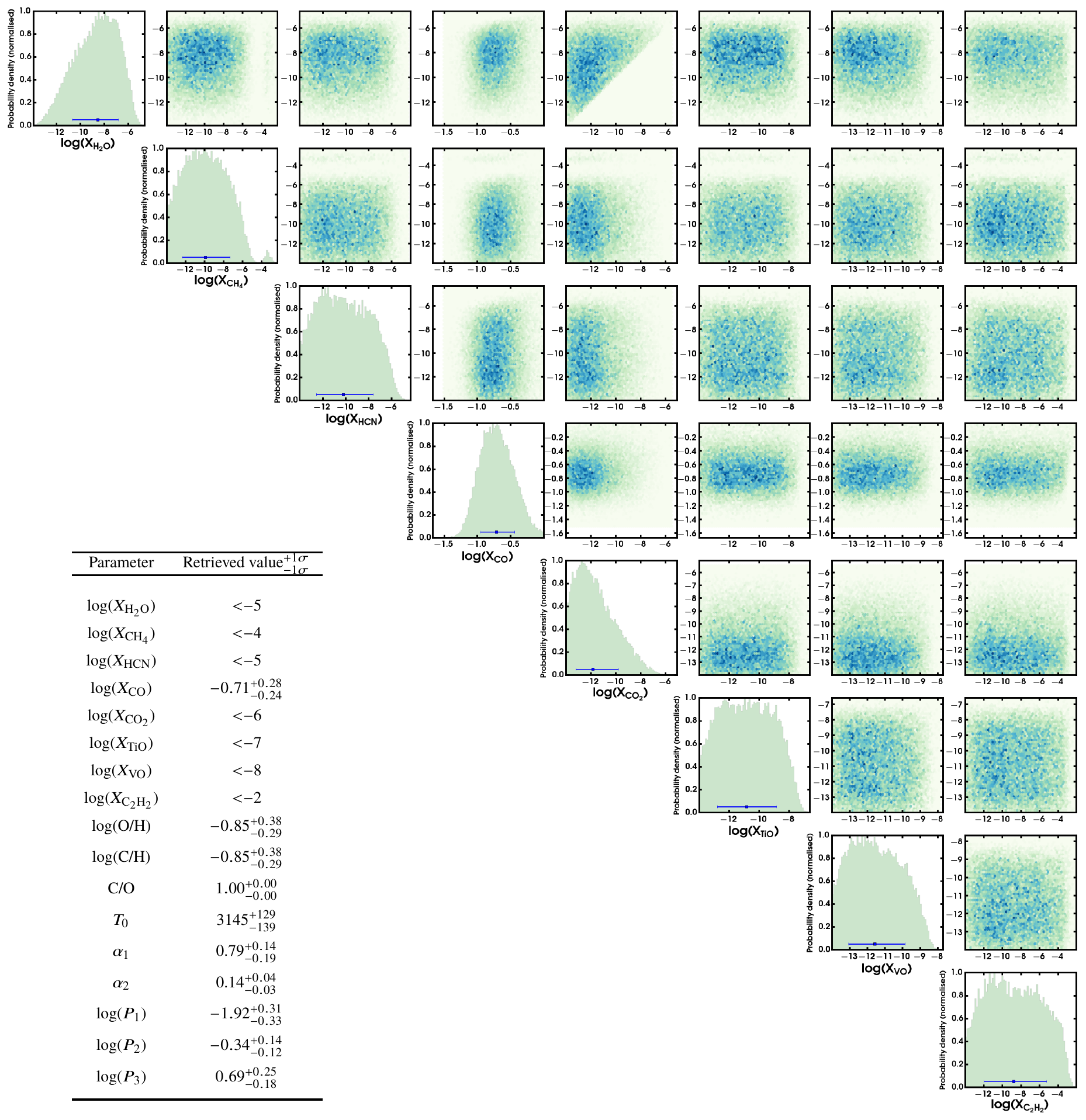}
\caption{Posterior distributions from our spectral retrieval. The mixing ratios are quoted as common log values. H$_2$O and CO$_2$ provide only upper limits, but the high CO abundance implies a high metallicity and high C/O.}
\label{fig:posterior}
}
\end{figure}

Figure~\ref{fig:posterior} shows the posterior probability distributions of all the model parameters. The data require a CO volume mixing ratio of $19^{+18}_{-8}$\% in the atmosphere, which is $380^{+360}_{-160}\times$ the amount expected for a solar abundance atmosphere at this temperature in thermochemical equilibrium. The high CO abundance is primarily constrained by the emission required to explain the 4.5\,$\mu$m IRAC point as well as the absorption trough in the WFC3 band at 1.6-1.7\,$\mu$m. We detect no other chemical species (see Figure~\ref{fig:posterior}). In particular, the non-detection of H$_2$O at both 1.4\,$\mu$m and 6\,$\mu$m provides a robust 3$\sigma$ upper-limit of $10^{-6}$ on the volume mixing ratio. The sum-total of constraints on the chemical species lead to a super-solar metallicity in the planet (C/H = O/H = $283^{+395}_{-138} \times$ solar O/H) and a C/O ratio of $\sim$1. 

We also conducted free-chemistry retrievals with no priors on the CO$_2$ abundance and find the same key results. For both cases, the data require a strong thermal inversion, a C/O ratio of $\sim$1, and a super-solar metallicity.

\section{Discussion}

The constraints on the chemical abundances are consistent with expectations for a high C/O ratio atmosphere in the high temperature regime of WASP-18b \citep{madhu2012, Moses:2013jh} where chemical equilibrium is expected to be satisfied. At high temperatures, H$_2$O is expected to be the most dominant oxygen-bearing molecule for a solar-abundance elemental composition (e.g. with a C/O = 0.5) \citep{madhu2012,Moses:2013jh}. In contrast, the low-abundance of H$_2$O observed is possible only if the overall metallicity and O abundance were low, or if the C/O ratio were high. Given the high abundance of CO we retrieve, the only plausible solution is both a high oxygen abundance and a high C/O ratio. The constraints on all the other species are also consistent with this scenario.  While we cannot rule out a contribution from CO$_2$ emission in the 4.5\,$\mu$m \textit{Spitzer} band, the high abundance of CO$_2$ needed would be chemically inconsistent with the non-detection of H$_2$O, and we therefore believe this scenario to be unlikely.

Our inferences for this planet indicate an unusual atmosphere in several respects, calling for comment on the reliability of our conclusions.  While the inference of a temperature inversion {\it per se} is no longer surprising for strongly irradiated planets \citep{Evans:2017dp, Haynes:2015cf}, both the very high metallicity and C/O $\sim 1$ have less precedent. Those aspects are forced upon us by the lack of observed water in the WFC3 and \textit{Spitzer} bandpasses, by the slight decrease at the long end of the WFC3 band, and by the \textit{Spitzer} photometry point at 4.5\,$\mu$m.  The non-detection of WFC3 water is certainly robust - several independent eclipses show no sign of the band head that should occur at 1.35\,$\mu$m (Figure~2). We reiterate that the inference of a thermal inversion hinges critically on the single Spitzer photometric point at 4.5\,$\mu$m. Previously, \citet{nymeyer} postulated a temperature inversion for exactly that reason. Since our eclipse depth agrees with those from previous analyses \citep{nymeyer, Maxted:2013ey}, we consider this measurement robust with regard to analysis technique. Nevertheless, the photometry does not reveal the resolved band structure of the 4.5\,$\mu$m CO band in emission that would lead to an unequivocal detection of molecular emission. However, given the data we have and the successful checks on our data analysis procedures, the unusual atmosphere of WASP-18b is a compelling conclusion. Our observations also reveal the first instance where both absorption and emission features are seen in the spectrum of an exoplanet, both due to CO. The absorption at $\sim$1.6 $\mu$m is caused by a weaker CO band compared to the emission in a stronger CO band in the 4.5 $\mu$m region. As shown by the contribution functions in Fig.~3, the 1.6 $\mu$m region in the spectrum probes the lower atmosphere due to the lower opacity compared to the 4.5 $\mu$m band which probes the upper atmosphere due to a higher opacity in that spectral region. Note that simultaneous absorption and emission in the same molecule is observed in the Earth's infrared spectrum, specifically in the 15\,$\mu$m band of CO$_2$, due to the temperature structure at the tropopause and stratosphere \citep{hanel}. 

If confirmed, the atmospheric properties of WASP-18b open a new regime in the phase space of hot Jupiters. Classically, thermal inversions in hot Jupiters were suggested to be caused by TiO and VO in very high temperature atmospheres \citep{Hubeny:2003eb,Fortney:2008p81}. All studies so far have focused on the plausibility of TiO/VO as a function of various parameters and processes such as settling and cold traps \citep{Spiegel:2009p7904}, stellar chromospheric emission \citep{Knutson:2010p6564}, C/O ratio \citep{madhu2011}, dynamics \citep{Parmentier:2013fd,menou2012}, etc.  For TiO/VO to be abundant enough to cause thermal inversions, the C/O balance must be approximately 0.5 or lower \citep{madhu2011}. Planets with high C/O ratios were not predicted to host thermal inversions since their TiO/VO abundances would be severely depleted \citep{madhu2012}; however, recent work suggests other processes, such as inefficient atmospheric cooling, could lead to an inverson \citep{Molliere:2015}. Alternatively, oxygen-poor absorbers may play a similar role to TiO and VO \citep{zahnle}. The two hot Jupiters for which thermal inversions have been detected have both showed signatures of TiO/VO in their atmospheres: WASP-33b \citep{Haynes:2015cf} and WASP-121 \citep{Evans:2017dp}. WASP-18b is the first system which shows a thermal inversion along with a high C/O ratio of $\sim$1 with no evidence for TiO/VO, and hence provides a new test case for theories of thermal inversions in hot Jupiters.

WASP-18b's unique atmospheric composition implies an interesting constraint for planetary formation theories. Its metal-enrichment is a factor of 1000 more than that predicted by the inverse mass-metallicity relationship for a 10$M_{J}$ planet \citep{Kreidberg:2014hi}. High metallicity and a C/O ratio of 1 are plausibly explained by formation from extremely CO-rich gas beyond the water condensation line \citep{Madhusudhan:2014} or upper atmospheric enrichment in carbon and oxygen due to ablation of icy planetesimals during late-stage accretion \citep{Pinhas:2016}. Future eclipse observations with the James Webb Space Telescope and improved modeling of giant planet accretion processes will help clarify the details of WASP-18b's formation history.
 
\acknowledgments
These observations are associated with HST program GO-13467 (PI. J.Bean) and Spitzer program GO-60185 (PI. P. Maxted). The authors thank Hannah Wakeford for assistance on data reduction algorithms and Eric Lopez for helpful discussions on interpretation. K. Sheppard, A. Mandell, P. Tamburo and D. Deming acknowledge funding from HST Program GO-14260 and a grant from the NASA Astrophysics Data Analysis Program. S. Gandhi acknowledges financial support from the Science and Technology Facilities Council (STFC), UK. A. Pinhas is grateful for research support from the Gates Cambridge Trust.

\clearpage

\clearpage

\begin{table*}
     
\label{table:spec}      
\centering       
\caption{Thermal Emission Spectrum}  

\begin{tabular}{c c |  c c}     
\hline\hline      
         
    Wavelength ($\mu$m) & Eclipse Depth (ppm) & Wavelength ($\mu$m) & Eclipse Depth (ppm) \\
\hline
    1.118-1.136 & 818 $\pm$ 28  &  1.434-1.452 & 1105 $\pm$ 25  \\  
    1.136-1.155 & 847 $\pm$ 26  &  1.452-1.471 & 1107 $\pm$ 25  \\
    1.155-1.173 & 858 $\pm$ 24  &  1.471-1.489 & 1088 $\pm$ 24 \\
    1.173-1.192 & 784 $\pm$ 25  &  1.489-1.508 & 1155 $\pm$ 28  \\
    1.192-1.211 & 944 $\pm$ 26  &  1.508-1.527 & 1159 $\pm$ 28 \\ 
    1.211-1.229 & 885 $\pm$ 26  &  1.527-1.545 & 1162 $\pm$ 28  \\
    1.229-1.248 & 913 $\pm$ 25  &  1.545-1.564 & 1077 $\pm$ 30  \\
    1.248-1.266 & 927 $\pm$ 25  &  1.564-1.582 & 1139 $\pm$ 30  \\
    1.266-1.285 & 900 $\pm$ 24  &  1.582-1.601 & 1130 $\pm$ 28  \\
    1.285-1.304 & 919 $\pm$ 25  &  1.601-1.620 & 1045 $\pm$ 34 \\
    1.304-1.322 & 957 $\pm$ 24  &  1.620-1.638 & 1019 $\pm$ 31  \\
    1.322-1.341 & 961 $\pm$ 23  &  1.638-1.657 & 1014 $\pm$ 38  \\
    1.341-1.359 & 1022 $\pm$ 25 &  2.15 & 1300 $\pm$ 300$^a$  \\ 
    1.359-1.378 & 1029 $\pm$ 29 &  3.6 & 2973 $\pm$ 70 \\
    1.378-1.396 & 1066 $\pm$ 26 &  4.5 & 3858 $\pm$ 113 \\
    1.396-1.415 & 1097 $\pm$ 25 &  5.8 & 3700 $\pm$ 300$^b$ \\
    1.415-1.434 & 1145 $\pm$ 25 &  8.0 & 4100 $\pm$ 200$^b$ \\
   
\hline
\end{tabular}
\tablecomments{WFC3 bin size =  0.0186$\mu$m}
\tablenotetext{1}{Zhou et al, 2015}
\tablenotetext{2}{ Nymeyer et al, 2011}
\end{table*}

\clearpage

\end{document}